\documentclass[10pt, superscriptaddress, prl, aps, twocolumn]{revtex4-2}

\usepackage{amssymb, amsmath}
\usepackage{graphicx}
\usepackage{physics, xcolor, orcidlink, fontawesome}
\usepackage[table]{xcolor}

\usepackage{algorithmicx}
\usepackage[noend]{algpseudocode}
\usepackage[most]{tcolorbox} 
\newtcolorbox{algobox}[1][]{colback=white, colframe=black, sharp corners, boxrule=0.8pt, left=6pt, right=6pt, top=6pt, bottom=6pt, enhanced}
\usepackage[normalem]{ulem}

\newcommand\scriplus{\ensuremath{\mathcal{I}^+}}

\newcommand{\IoA}{Institute of Astronomy, University of Cambridge, Madingley Road, Cambridge, CB3 0HA, UK}
\newcommand{\DAMTP}{Department of Applied Mathematics and Theoretical Physics, Centre for Mathematical Sciences, University of Cambridge, Wilberforce Road, Cambridge, CB3 0WA, UK}
\newcommand{\KICC}{Kavli Institute for Cosmology, University of Cambridge, Madingley Road, Cambridge, CB3 0HA, UK}

\begin{document}

\title{Quasinormal mode content of binary black hole ringdowns} 

\author{Richard Dyer\,\orcidlink{0009-0008-3720-6092}}
\email{richard.dyer@ast.cam.ac.uk} 
\affiliation{\IoA} 

\author{Christopher J.\ Moore\,\orcidlink{0000-0002-2527-0213}}
\email{christopher.moore@ast.cam.ac.uk}
\affiliation{\IoA} 
\affiliation{\DAMTP} 
\affiliation{\KICC}

\date{\today}

\begin{abstract}
    We present a fully Bayesian, data-driven framework for identifying quasinormal modes in high-accuracy Cauchy–Characteristic Evolution (CCE) gravitational waveforms. 
    Applying this to a public catalog, we identify QNM overtones, retrograde modes, and nonlinear modes up to cubic order in the ringdown.
    Multiple high-order overtones are found near the merger, confirming their physical significance for modeling the ringdown. 
    The ringdown mode content is tabulated across a wide range of start times for all available simulations, providing a systematic reference for theoretical and observational studies.
    We also search for late-time power-law tails, which are, as expected, absent from the CCE waveforms.
\end{abstract}

\maketitle

{\textit{\textbf{Introduction}}---}Following a binary black hole (BH) merger, the remnant emits gravitational waves (GWs) as it settles into its final stationary state. 
This process, the ringdown, is well described as a perturbation of the final BH.
Compared to the earlier inspiral and merger, the simpler, perturbative dynamics of the ringdown makes it an attractive arena for 
testing general relativity \cite{Carullo:2025oms, 2025arXiv250523895B} through the nature and uniqueness of the Kerr solution \cite{2004CQGra..21..787D, 2019PhRvL.123k1102I, 2025arXiv250908099T} and the horizon area law of BH thermodynamics \cite{2021PhRvL.127a1103I, 2024PhRvD.110d4018C, 2025PhRvL.135k1403A} with strong gravitational fields. 

Quasinormal modes (QNMs), damped sinusoidal oscillations at specific frequencies, are prominent in the ringdown \cite{2009CQGra..26p3001B}. 
Ringdown tests of general relativity generally require the detection of multiple QNMs. 
The fundamental quadrupolar QNM is usually the loudest and is routinely found in GW observations \cite{2021arXiv211206861T}.
However, other modes can dominate and the relative importance among the subdominant modes is less clear and depends on the properties of the progenitor binary \cite{2012PhRvL.109n1102K, 2024PhRvD.110j3037P, 2025PhRvD.112b4077M, 2025PhRvD.111f4052Z}.
Observational ringdown studies need guidance from theory to know which subdominant modes to target on an event-by-event basis.

The only reliable theoretical tool for modeling the merger is numerical relativity. 
In particular, for high-fidelity GW ringdown signals, we use SpECTRE Cauchy Characteristic Evolution (CCE) numerical simulations \cite{2020PhRvD.102d4052M, 2021arXiv211008635M}, which include an exterior solution based on null (or characteristic) slices that extend all the way to future null infinity ($\scriplus$) where the GW signal is defined.

Our task is to determine the QNM content of a numerical waveform.
This is challenging due in part to the ambiguity in the ringdown start time, the presence of nonlinearities not described by perturbation theory, and the risk of overfitting with many QNMs; see, for example, \cite{2018PhRvD..97j4065B, 2020arXiv200400671O, 2019PhRvX...9d1060G}. 

Previous attempts have largely relied on least-squares fits to individual waveform harmonics \cite{PhysRevD.102.084052, 2019PhRvX...9d1060G, 2021PhRvD.103j4048D, 2021PhRvD.104l4072F, 2021PhRvD.103h4048F, 2024PhRvD.109d4069C, 2025PhRvD.111h4041G, 2025PhRvD.112f4016M}. 
The quality of the fits is assessed using a simple mismatch and the above difficulties are handled with \emph{ad hoc} approaches such as free-frequency fits or amplitude-stability checks across start times. 
Rational filters offer an alternative approach but do not measure the amplitudes of the QNMs and have not been widely adopted \cite{2022PhRvD.106h4036M}. 
A few Bayesian approaches to studying the time dependence of the overtone amplitudes have also been explored \cite{2024PhRvD.109l4030C, 2024PhRvD.109j1503R, 2024JCAP...10..061C}.

This letter describes a new Bayesian method for determining the mode content of the ringdown. 
Instead of least-squares fits, posterior distributions on the model parameters are obtained using a likelihood function that accounts for the uncertainty in the numerical waveforms \cite{methods_paper}. 
The mode content is determined independently at each start time and overfitting is naturally avoided by the use of evidence ratios (Bayes factors) to determine which modes to include and which penalizes overly complex models. 
Posterior predictive checks (PPCs) are used to assess the fit quality.

{\textit{\textbf{Theory}}---}The GW signal is conventionally decomposed into spin-weighted spherical harmonics (indexed by $\beta\!=\!(\ell,m)$).
The QNM model for the signal in each harmonic is a sum of damped sinusoids and the overall amplitude decays inversely with distance $r$ from the BH:
\begin{equation}
    h_{\rm QNM}^{\beta} (t) = \frac{M}{r} \sum_{\alpha} C_\alpha \mu_{\alpha}^{\beta} e^{-i \omega_\alpha(t-t_0)} . \nonumber
\end{equation}
The QNMs (indexed by $\alpha\!=\!(\ell, m, n, p)$) have amplitudes $C_\alpha$ that are regarded as free parameters.
The QNM frequencies $\omega_\alpha$ and spherical-spheroidal mode-mixing coefficients $\mu_\alpha^\beta$ are known from BH perturbation theory and are functions of the remnant BH mass $M_f$ and spin $\chi_f$, calculated using Ref.~\cite{Stein:2019mop}. 
The polar, azimuthal and radial indices vary in the ranges $\ell\!\geq\!2$, $|m|\!\leq\!\ell$, $n\!\geq\!0$, and the parity $p\!\in\!\{+,-\}$.
QNMs with $\mathrm{sign}(p)\!=\!\mathrm{sign}(m)$ co-rotate with the remnant (prograde); QNMs with opposite signs are retrograde.
The model is only used after the start of the ringdown, $t\!\geq\!t_0$.
Fitting all the spherical harmonics $\beta$ of the GW signal simultaneously effectively performs an angle-averaged fit over the sphere at \scriplus.
The mass scale $M$ is included to make the strain amplitudes $C_\alpha$ dimensionless. 
(We use natural units, $G\!=\!c\!=\!1$.)

The model can be extended by including additional terms.
Second-order perturbation theory predicts the existence of quadratic QNMs associated with pairs $(\alpha, \alpha')$ of linear QNMs with frequencies $\omega_{\alpha\alpha'}\!=\!\omega_{\alpha}\!+\!\omega_{\alpha'}$ \cite{2023PhRvD.107d4040L}. 
Second-order perturbation theory does not make a simple prediction for the angular structure of the quadratic QNMs \cite{2024PhRvD.109j4070M}.
Nevertheless, it is possible define an analog of the mode-mixing coefficients to describe the shape of the quadratic QNMs to better than one part in $10^3$ \cite{2025PhRvD.111b4002D} 
(`prediction C' of Ref.~\cite{2025PhRvD.111b4002D} is used for the results shown here, which are also compared with those found using predictions Bi, Bii, and D).
Higher-order cubic QNMs, associated with triples $(\alpha, \alpha', \alpha'')$ of linear QNMs, can also be included in a similar way. 

Previous attempts to model the ringdown have found it necessary to include constant offsets for each harmonic in the model \cite{2024PhRvD.109d4069C, 2025PhRvD.111h4041G}. 
The need for these is reduced when fitting to the Bondi news ($\mathcal{N}\!=\!\partial_t h$) instead of directly to the strain. 
Nevertheless, constant offsets are included in the model here even though they are rarely needed.

Linear perturbation theory also predicts power-law tails (or `Price' tails) at late times \cite{1972PhRvD...5.2419P}. 
These can also be included in the model as described below. 

Our final ringdown model is a sum of all these types of term. 
Each term is referred to as a mode.
Exactly which modes should be included is determined by the Bayesian ringdown model building algorithm described below.

{\textit{\textbf{Numerical Waveforms}}---}We use simulations from the public SpECTRE CCE catalog \cite{SXS_CCE_catalog, spectrecode}. 
These are evolved to \scriplus{} using CCE \cite{2020PhRvD.102d4052M, 2021arXiv211008635M}. 
The waveforms are transformed into the superrest frame 
\cite{2022PhRvD.105j4015Z, 2022PhRvD.106h4029M, 2024CQGra..41v3001M, Boyle2013, BoyleEtAl:2014, Boyle2015a, mike_boyle_2020_4041972, sxs_software} 
and post-processed as described in Ref.~\cite{methods_paper}.
The time is shifted such that $t\!=\!0$ is the peak of the strain in the $\beta\!=\!(2,2)$ harmonic.

The simulations are available at several numerical resolutions.
The difference between the highest two resolutions is taken as a conservative estimate of the numerical uncertainty.
Furthermore, because these waveforms were produced using similar setups, their uncertainties are expected to have similar properties across the catalog.
Ref.~\cite{methods_paper} exploited this by introducing a physically-motivated Gaussian process model for the uncertainty, trained on the catalog to pool information among all available waveforms. 
The trained Gaussian process was used to construct a likelihood function for the ringdown model.

{\textit{\textbf{Ringdown Model Building}}---}This section describes a fully Bayesian, data-driven method for determining the ringdown mode content. 
Modes are selected based on their Bayes factors and the issue of when the resulting model gives an adequate fit to the data (i.e.\ over- or under-fitting) is addressed using a PPC.

Start by picking a simulation from the catalog, a subset of the spherical harmonics to model, and a start time $t_0$ for the ringdown.
Initially, consider a ringdown model that contains no modes at all; the set of detected modes in the model is empty, $\mathcal{D}\!=\!\emptyset$, and the corresponding signal model is simply $h^\beta(t)\!=\!0$.
The algorithm also starts with a set of candidate modes, $\mathcal{C}$; this is initialized with all the modes to be considered for inclusion (typically, around $\sim 100$) and includes all the mode types discussed above (except the tails which are handled separately).

For each candidate $c\in\mathcal{C}$ the numerical waveform harmonics are fit in the range $t_0\!\leq\!t\!<\!t_0 +100M$ with a ringdown model containing the previously detected modes \emph{and} $c$; i.e.\ $\mathcal{D}\!\cup\!\{c\}$. 
The fit is done using the computationally efficient Bayesian methodology described in Ref.~\cite{methods_paper}, with flat priors on the real and imaginary parts of the amplitudes, and with the remnant mass and spin fixed to the numerical relativity values.
The fits were performed to the Bondi news but our results are quoted converted into GW strain. 
The significance of the candidate, $\mathcal{S}_c$, is calculated; this is related to the Bayes factor between ringdown models with mode contents $\mathcal{D}$ and $\mathcal{D}\!\cup\!\{c\}$ with equal prior odds \cite{methods_paper}. 
This is repeated for all candidates and the one with the highest significance, $c'$, is added to the detected modes; $\mathcal{D}\!\gets\!\mathcal{D}\!\cup\!\{c'\}$.
It is also removed from the candidates; $\mathcal{C}\!\gets\!\mathcal{C}\!\setminus\!\{c'\}$.
This process of adding the most significant candidate one at a time is repeated until either no candidates remain or none exceeds a threshold $\mathcal{S}_{\rm threshold}=0.9999$ corresponding approximately to measuring a nonzero mode amplitude at $\sim 4\sigma$ significance.

In practice, this iterative process of adding modes can be sped up by initializing $\mathcal{D}$ with modes certain to be present (e.g.\ the fundamental linear QNMs) and by only checking a restricted set of candidates at each iteration.
This is described in more detail in the supplement where a step-by-step example of the algorithm is also presented.

\begin{figure*}[t]
    \centering
    \includegraphics[width=0.99\textwidth]{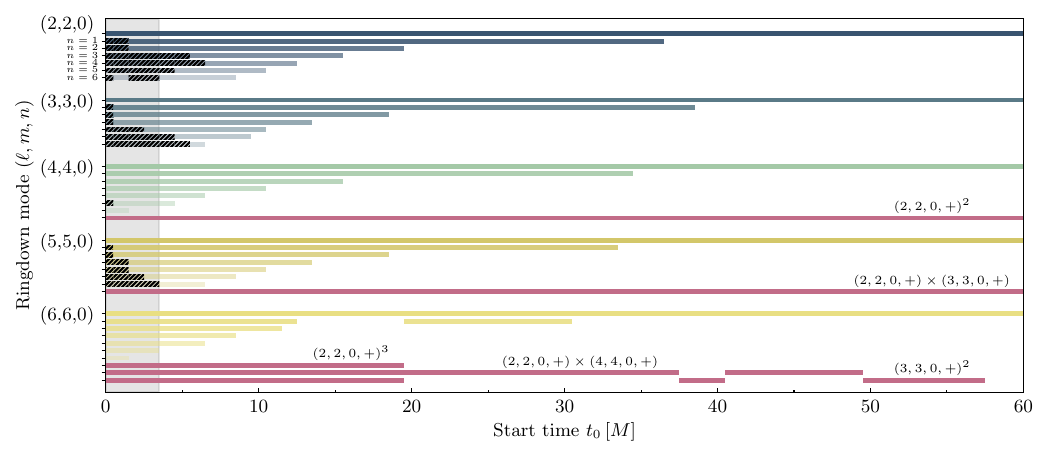}
    \caption{ \label{fig:mode_content}
        The ringdown mode content for simulation 0010.
        Fits are performed independently at each start time, $t_0$.
        Colored horizontal bands indicate a mode detected at that $t_0$.
        Prograde QNMs are grouped and colored based on their $(\ell,m)$ indices with the fundamental ($n\!=\!0$) uppermost and overtones ($n\!\geq\!1$) below.
        If the retrograde mode is also detected then the band is hatched.
        Nonlinear QNMs are shown with the harmonic that they mix strongly with.
        The shaded region $t_0\!\leq\!3M$ indicates where the PPC fails and the model fits the data poorly.
        Similar plots for all simulations in the catalog are available at \href{https://bgp-qnm-fits.github.io/bgp_qnm_content/}{\faLink} \cite{website}.
    }
\end{figure*}

This process produces a set of modes $\mathcal{D}$ detected at high significance.
However, this alone does not necessarily imply that a ringdown model with these modes is a good fit; for example, at early times no ringdown model fits the data well.
For this reason, the final stage of the algorithm involves a PPC using a ringdown model with the modes in $\mathcal{D}$.
The amplitude ($L^2$ norm) of the fit residuals is compared with the expected generalized $\chi^2$ sampling distribution under the assumed Gaussian process model for the numerical uncertainty (see, for example, Fig. 10 of Ref.~\cite{methods_paper}). This statistical null test is made possible by the new Gaussian process model which provides an improved description of the numerical noise. This test of the quality of fit prevents over- and/or under-fitting and is a vital part of the algorithm.
The cumulative distribution function value of the fit residuals with respect to this distribution gives a measure of fit quality, with large values ($>\!0.9$) indicating a poor quality fit.

This letter reports results from fits to all 13 numerical simulations in the public catalog across a wide range of ringdown start times.
For each simulation, we are primarily interested in the mode content of the loudest harmonic at each value of $\ell$ (up to an $\ell_{\rm max}$ such that $\geq\!99\%$ of the radiated GW energy is contained in the harmonics $\ell\!\leq\!\ell_{\rm max}$).
For the non-spinning mass ratio $1\!:\!4$ simulation 0010, these target harmonics are $(2,2)$, $(3,3)$, $\ldots$ $(6,6)$; for non-precessing systems, the corresponding $m\!<\!0$ harmonics are omitted because they are related by symmetry.
We find that in order to reliably model the ringdown mode content of the $(\ell,m)$ harmonic it is necessary to include $(\ell+1,m)$; this is due to the strong ($\mu\!\sim\!0.1$, depending on the remnant BH spin) mode-mixing.
Therefore, for simulation 0010, we also include the $(3,2)$, $(4,3)$, $\ldots$, $(7,6)$ harmonics in our modeling, although we do not report the mode contents for these.
For precessing binaries with misaligned component BH spins the $m\!<\!0$ harmonics were also included.
For each simulation, the included $\beta$ harmonics are listed in table~\ref{tab:spherical_harmonics} of the supplement.

The set of candidates $\mathcal{C}$ includes the linear QNMs $\alpha\!=\!(\ell, m, n, p)$ with $\ell, m$ indices matching the included $\beta$ harmonics, with overtones $0\leq\!n\!\leq 6$ and retrograde modes $p\!\in\!\{+,-\}$.
The set $\mathcal{C}$ also included constant offsets for each of the included harmonics and the nonlinear QNMs formed from combinations of linear fundamental ($n\!=\!0$) QNMs with $\ell, m$ indices in the included $\beta$ modes expected to contribute strongly to the target harmonics;
these are the $(2,2,0,+)^2$, $(3,3,0,+)\!\times\!(2,2,0,+)$, $(3,3,0,+)^2$, and $(4,4,0,+)\!\times\!(2,2,0,+)$ quadratic QNMs and the $(2,2,0,+)^3$ cubic QNM (and corresponding prograde modes with $m\!<\!0$).
Nonlinear QNMs are only considered as candidate when all their sourcing linear QNMs are already in $\mathcal{D}$.
The tails are treated separately, as described below.

{\textit{\textbf{Results and Discussion}}---}As an example of our results, the ringdown mode content for the target harmonics of the mass ratio $1\!:\!4$ simulation 0010 is shown in Fig.~\ref{fig:mode_content}.
Results for the whole catalog are at \href{https://bgp-qnm-fits.github.io/bgp_qnm_content/}{\faLink} \cite{website}.

At early start times, $t_0\!\lesssim\!9M$, the algorithm identifies overtones up to $n\!=\!6$ in multiple spherical harmonics.
For $t_0\!\gtrsim\!4M$ the ringdown model gives a good fit to the GW signal as judged by the PPC. 
As the start time increases, the overtones drop out of the model in a predictable way. 
The only exception is the $\alpha\!=\!(6,6,1,+)$ mode, which reappears again around $t_0\!\approx\!20M$.
This behavior has been observed previously \cite{methods_paper} and is associated with QNM amplitudes over time deviating from the expected spiral pattern in the complex plane. The fact that our data-driven, Bayesian approach confidently identifies multiple high-order overtones at the times where they would theoretically be expected, adds to the evidence that they play an important role in the ringdown.

\begin{figure*}[t]
    \centering
    \includegraphics[width=.99\textwidth]{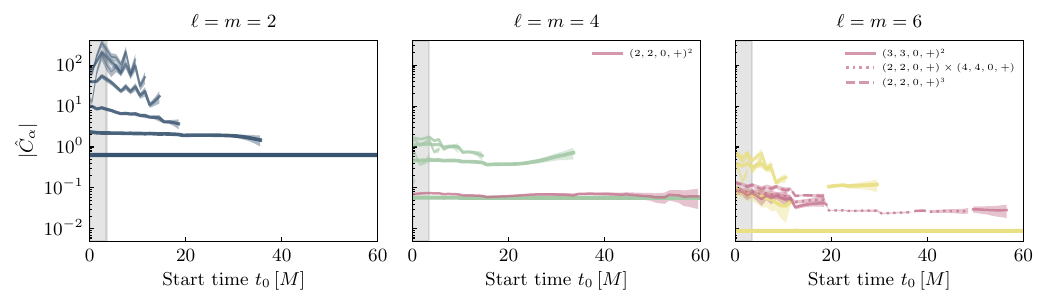}
    \caption{ \label{fig:amp_stability}
        Plots of selected (decay-corrected) amplitudes $\hat{C}_\alpha$ \cite{methods_paper} for the detected modes in simulation 0010 as a function of $t_0$.        
        Horizontal lines indicate modes that decay exponentially at the rate predicted by perturbation theory.
        The left/center/right plots show the $(\ell,m)\!=\!(2,2)$/$(4,4)$/$(6,6)$ sectors from Fig.~\ref{fig:mode_content} with matching colors. 
        Thick lines denote fundamental QNMs with thinner lines denoting higher overtones; nonlinear QNMs are indicated in the legends (only prograde modes are shown).
        The central lines indicate the median posterior amplitudes and the shaded regions indicate the $90\%$ ranges; in most cases the shaded regions are too small see clearly at this scale.
        Similar plots for all simulations in the catalog are available at \href{https://bgp-qnm-fits.github.io/bgp_qnm_content/}{\faLink} \cite{website}.
    }
\end{figure*}

Nonlinear QNMs are also identified. 
The dominant quadratic QNMs appear clearly in both the $\beta\!=\!(4,4)$ and $(5,5)$ sectors, with these modes persisting for longer than any overtone due to the slower decay rate of the fundamental linear QNMs that source these perturbations.
This is consistent with earlier results obtained using least-squares fits \cite{2023PhRvL.130h1402M, 2023PhRvL.130h1401C}. 
Three nonlinear QNMs, including the $(2,2,0,+)^3$ cubic QNM, are identified in the $\beta\!=\!(6,6)$ harmonic for a range of start times.
The $(6,6)$ is the quietest target harmonic, making it the hardest to model accurately, especially at late times.
The model-building algorithm struggles at late times; at start times $t_0\!\gtrsim\!37M$, while it consistently finds support for some mode in addition to the fundamental $(6,6,0,+)$, it alternates its preference between the $(3,3,0,+)^2$ and $(2,2,0,+)\!\times\!(4,4,0,+)$ quadratic QNMs.
By contrast, the cubic QNM is consistently identified with a stable amplitude (varying by $\lesssim\!35\% $) for all $t_0\!\lesssim\!20M$. 
We also find these nonlinear QNMs in all of the other mass ratio $1\!:\!4$ simulations where the $\beta\!=\!(6,6)$ harmonic was analyzed, 0010-0013, although the results of the extreme precessing system 0013 are of noticeably poorer quality.

The results for the detected nonlinear QNMs are robust to changes in how the shapes of these modes are modeled.
The mode content was also determined using predictions Bi, Bii, and D \cite{2025PhRvD.111b4002D} for their mode-mixing coefficients with similar results \href{https://bgp-qnm-fits.github.io/bgp_qnm_content/sim_0010.html#nonlinear-mode-mixing-alternatives}{\faLink} \cite{website}.

For simulation 0010, Fig.\ref{fig:mode_content}, retrograde QNMs are only occasionally identified by the algorithm at very early start times.
At these early times the overall ringdown model mostly does not give a good fit to the numerical relativity data, as indicated by the PPC.
However, retrograde modes are found more prominently in other simulations in the catalog; notably in the $\beta\!=\!(3,3)$ harmonic of the equal-mass systems with anti-aligned component spins, 0005-0007.

It is possible to examine how the amplitudes of the modes in these fits change with ringdown start time.
We stress that the model-building algorithm fits each $t_0$ independently, with no requirement that either the mode content or amplitudes vary continuously with time.
Nevertheless, the Bayesian fits do correctly infer the expected decay rates for the fundamental linear QNMs, many low-order overtones, and nonlinear QNMs; see Fig.~\ref{fig:amp_stability}.

The high-order overtone amplitudes are less stable with varying $t_0$.
This behavior is at the root of the controversy over the physical significance of QNM overtones \cite{2019PhRvX...9d1060G, 2021PhRvD.103d4054M, 2021PhRvD.104l4072F, 2023PhRvD.108j4020B, 2023PhRvD.108d4032N, 2024PhRvD.109d4069C, 2024PhRvD.109j4070M, 2025PhRvD.111j4043O, 2025PhRvD.111h4041G}. 
For a fixed start time, data-driven detection criteria show a clear preference for the inclusion of overtones; this is shown here using a Bayesian approach but has also been found by many authors using least-squares fits.
However, if the fits are performed at different start times, and/or if the frequencies of the overtones are allowed to vary, and consistency with perturbation theory predictions is used to define an \emph{ad hoc} physically-motivated detection criterion, then overtones are more likely to be missed. 
The frequencies of the overtones are not expected to be well measured because of their short decay times, and their amplitudes at early times might not be expected to be stable due to confusion with the ringdown prompt response \cite{2025arXiv250909165O}. 

A consistency test of the detected mode content is to re-perform the Bayesian fits with the remnant mass and spin allowed to vary as free parameters. The resulting posteriors are expected to be consistent with the numerical values, this can be quantified using the squared distance in the `Regge plane', $\epsilon^2\!=\!(\Delta M_f / M_f)^2\!+\!\Delta \chi_f^2$. This was done for two ringdown models: with the detected modes $\mathcal{D}$ and, for comparison, with the candidate modes $\mathcal{C}$. Root-mean-square $\epsilon$ values of $\mathcal{O}(10^{-4})$ are found across a range of start times for all simulations in the catalog \href{https://bgp-qnm-fits.github.io/bgp_qnm_content/sim_0010.html#epsilon}{\faLink} \cite{website}. 

In addition to QNMs, the model building algorithm also considered adding constant offset modes to the model. 
When fitting to the Bondi news (as was done here) which has more regular behavior at late times than the GW strain, these constant offsets are usually not detected by the algorithm.

The Bayesian approach to identifying modes in the ringdown can also be adapted to search for tails.
The model is extended by including a term proportional to $A^\beta (t-t_\beta)^{-\lambda_\beta}$ in one spherical harmonic at a time (no mode-mixing is modeled for the tails).
The model now depends on the QNM, nonlinear QNM and constant offset parameters described above (collectively denoted $\vartheta$) and the tail parameters: $A^{\beta}$, $\lambda_{\beta}$ and a reference time $t_{\beta}$.
Because the model is linear in the $\vartheta$ parameters (see Ref.~\cite{methods_paper}), these can be marginalized out analytically leaving a posterior on just the tail parameters. 
Sampling this posterior effectively marginalizes over the oscillatory parts of the ringdown allowing us to efficiently search for the quieter monotonically decaying tails.  

This tail posterior was sampled using \texttt{emcee} \cite{2013PASP..125..306F}. 
Because the tails are expected to dominate only at late times \cite{2024arXiv241206906M} an analysis was performed for $t_0 \!= \!50M$. 
The oscillatory mode content was fixed using the modes $\mathcal{D}$ detected by our algorithm (see, e.g.\ Fig.~\ref{fig:mode_content}). 
Further details of the tail analysis are given in the supplement.

Tails were searched for but not found in all of the target spherical harmonics of simulation 0010.
The posteriors on the tail amplitudes $\hat{A}^\beta$ at $t_0\!=\!50M$ were consistent with zero and place 90\% upper bounds of $|\hat{A}^{(2,2)}|\!<\!1.7 \times 10^{-5}$ \href{https://bgp-qnm-fits.github.io/bgp_qnm_content/sim_0010.html#power-law-tails}{\faLink} \cite{website}. 
Tighter upper bounds were found in the higher harmonics, $|\hat{A}^{(6,6)}|\!<\!3.6 \times 10^{-6}$. 
The absence of tails is expected because of the limited size of the Cauchy region in the CCE simulations and the absence of back reaction from outside this region on the binary. 
Tails have previously been identified only in Cauchy Characteristic Matching (CCM) simulations \cite{2024arXiv241206906M} which include back reaction, or in Cauchy simulations of head-on collisions with very large domains \cite{2024arXiv241206887D}. 
However, the Bayesian algorithm described here can be used to simultaneously identify QNMs and tails when catalogs of CCM simulations, or simulations using hyperboloidal coordinates, \cite{2024PhRvD.110l4033P} become available.

{\textit{\textbf{Acknowledgments}}---}We are grateful to E.\ Finch, M.\ Isi and K.\ Mitman for helpful discussions. 
We would also like to thank the anonymous reviewers for their constructive reports which helped improve this manuscript. R. D. acknowledges support from the Science \& Technologies Facilities Council
toward their Ph.D. studies (Grant No. ST/Y509139/1).

{\textit{\textbf{Data Availability}}---}The code to perform the Bayesian fits \cite{methods_paper} is available at \href{https://github.com/BGP-QNM-FITS/bgp_qnm_fits}{\faGithub}. 
The data and plotting scripts for the results described in this letter are available at \href{https://github.com/BGP-QNM-FITS/bgp_qnm_content}{\faGithub} and the plots themselves at the website \href{https://bgp-qnm-fits.github.io/bgp_qnm_content/}{\faLink} \cite{website}. 
We are grateful to the SXS collaboration for making the numerical relativity data publicly available \cite{SXS_CCE_catalog}.

\bibliographystyle{apsrev4-2}
\bibliography{references}

\clearpage
\onecolumngrid
\newpage
\begin{center}
  \textbf{\large{Supplemental Material}} \\
\end{center}
\twocolumngrid

\setcounter{equation}{0}
\setcounter{figure}{0}
\setcounter{table}{0}
\setcounter{page}{1}
\makeatletter
\renewcommand{\theequation}{S\arabic{equation}}
\renewcommand{\thefigure}{S\arabic{figure}}
\renewcommand{\thetable}{S\arabic{table}}

\section{Description of the ringdown model building algorithm}

\renewcommand{\figurename}{Algorithm}
\begin{figure}[t]
    \begin{algobox}
        \begin{algorithmic}[1]
            \State $\mathcal{C}\gets\begin{Bmatrix} (2,2,0,+), (2,2,1,+), \\ (3,2,0,+), (4,4,1,+), \\ (2,2,0,+,2,2,0,+), \ldots \end{Bmatrix}$  \Comment{Candidate Modes}
            \State $\mathcal{D}\gets\emptyset$  \Comment{Detected modes; initially empty}
            \State $S_{\rm max} \gets 1$
            \While{$S_{\rm max}>\mathrm{S}_{\rm threshold}$ \textbf{and} $|\mathcal{C}|\geq 1$} \Comment{Loop until no candidate exceeds threshold}
                \State $S_{\rm max} \gets 0$
                \For{$c\in\mathcal{C}$}  \Comment{Loop over candidates}
                    \State $\mathrm{modes}\gets\mathcal{D}\cup\{c\}$ 
                    \State $\mathrm{fit}\gets\mathrm{bayes\_fit}(\mathrm{modes})$ \Comment{Fit using candidate}
                    \State $S\gets\mathrm{sig} (\mathrm{fit}, c)$ \Comment{Significance of candidate}
                    \If{$S>S_{\rm max}$}
                        \State $S_{\rm max}\gets S$ 
                        \State $c'\gets c$ \Comment{Keep most significant mode $c'$}
                    \EndIf
                \EndFor
                \If{$S_{\rm max}>\mathrm{S}_{\rm threshold}$}
                    \State $\mathcal{D}\gets\mathcal{D}\cup\{c'\}$ \Comment{Add $c'$ to detected modes}
                    \State $\mathcal{C}\gets \mathcal{C}\setminus\{c'\}$ \Comment{Remove $c'$ from candidates}
                \EndIf
            \EndWhile
            \State \textbf{return} $\mathcal{D}$ \Comment{The final QNM content}
        \end{algorithmic} 
    \end{algobox}
    \caption{\label{alg:algorithm}
        Bayesian algorithm for determining the QNM content of the ringdown.
        The \texttt{bayes\_fit} (line 8) refers to a full Bayesian inference of a ringdown model with the specified mode content to the numerical data while \texttt{sig} (line 9) refers to the calculation of the significance of a particular mode $c$ (related to the Bayesian evidence) in that fit.
        In practice, the modifications described in the text speed this up.
    }
\end{figure}
\renewcommand{\figurename}{Fig.}

The Bayesian method for determining the content of the ringdown signal was described in the main text.
It is illustrated here using pseudocode in algorithm \ref{alg:algorithm}.

Implementing this directly can be computationally expensive when the number of candidates in $\mathcal{C}$ is large. 
For example, for simulation 0013 the number of candidates was $|\mathcal{C}|\!=\!310$.
At each iteration of the algorithm a Bayesian inference run and significance calculation must be performed for each candidate, and this must be repeated for every $t_0$.
To reduce the computational cost, we make the following modifications to algorithm \ref{alg:algorithm}.

First, the set of detected modes $\mathcal{D}$ is initialized with the fundamental prograde linear QNMs allowed by symmetry (i.e.\ $\alpha\!=\!(\ell, m, 0, \mathrm{sign}(m))$ with $\ell, m$ indices matching the included $\beta$ harmonics) which are always expected to be present.
On line 2, replace the empty initialization with `$\mathcal{D}\!\gets\!\{(\ell,m,0,+)\,|\,$for each $(\ell,m)$ numerical waveform mode to be fit$\}$'.
These fundamental prograde QNMs are also removed from the initial set $\mathcal{C}$.
This reduces the number of iterations of the algorithm.

Second, at each iteration we consider only a subset of candidates.
This subset, $\mathcal{C}'$, includes only the next prograde overtone for each spherical harmonic; for example, if $\mathcal{D}$ already contains $(\ell,m,0,+)$ and $(\ell,m,1,+)$, then $\mathcal{C}'$ includes $(\ell,m,2,+)$ but no higher overtones $(\ell,m,n\geq 3,+)$.
(All the retrograde overtones are tested at each iteration.)
This restriction is justified because the higher overtones are expected to decay faster and hence have lower significances, especially at later times.
The subset $\mathcal{C}'$ only includes nonlinear modes whose corresponding linear modes are already detected; e.g.\ the $(3,3,0,+)^2$ quadratic QNM is only considered as a candidate if the $(3,3,0,+)$ linear QNM is already in $\mathcal{D}$.
The subset $\mathcal{C}'$ also includes all the constant-offset modes.
On line 6, we replace the loop over $\mathcal{C}$ with `$\mathbf{for}\; c\in\mathcal{C'}\; \mathbf{do}$'. 
This greatly reduces the number of modes checked at each iteration.

These adjustments usually have no effect on the final detected mode content. 
This is demonstrated with a specific example in the next section.

\section{An Example of the Ringdown Model Building Algorithm} 

To illustrate the procedure, we show in a specific case how, and in what order, modes are added to the model.

For this example, the target spherical harmonics ${\beta\!\in\!\{(2,2), (3,2)\}}$ of the equal-mass simulation $0001$ are modeled from a ringdown start time of $t_0\!=\!20M$.
As described in the main text, accurate modeling of $(3,2)$ requires including the $(4,2)$ harmonic.
Therefore, the included harmonics are $\beta\!\in\!\{(2,2), (3,2), (4,2)\}$ although a reliable mode content is obtained only for the two target harmonics.
The candidate modes $\mathcal{C}$ included the linear QNMs $(2,2,n,\pm)$, $(3,2,n,\pm)$, and $(4,2,n,\pm)$ with $0\leq n\leq 6$ and a constant offset mode in each harmonic.

The algorithm \ref{alg:algorithm} was run \emph{without} the performance modifications described above.
Table~\ref{tab:qnm_mode_adding} shows the set of detected modes $\mathcal{D}$ in the order in which they were added and the significance of each mode at the point it was added to the model. 
The final three columns show how the fit improves as modes are added, quantified using the white-noise mismatch (Eq.~29 of Ref.~\cite{methods_paper}) in each harmonic.

As seen in table~\ref{tab:qnm_mode_adding}, for each $\beta$ the fundamental ($n\!=\!0$) QNM is identified first, followed by overtones in increasing $n$. 
Therefore, exactly the same set $\mathcal{D}$ is obtained if the algorithm is run with the performance modifications described above.

\section{Spherical Harmonics} 

Table~\ref{tab:spherical_harmonics} lists the dominant spherical harmonics used for each numerical relativity simulation in the catalog.

\begin{table*}[t]
    \caption{ \label{tab:qnm_mode_adding} 
        Table showing the ringdown mode content for an example fit to simulation $0001$ from $t_0=20M$ using harmonics $\beta\!=\!(2,2)$, $(3,2), (4,2)$. 
        At each iteration a mode $\alpha'$ is added to model (if the significance $\mathcal{S}_{\alpha'}$ exceeds the threshold). 
        The order in which modes are added is related to their importance in the model.
        As modes are added the quality of fit (quantified by the mismatches $\mathcal{M}_\beta$ in each harmonic) steadily improves.
        The total number of detected modes in the target harmonics is 5 but the total overall is $|\mathcal{D}|\!=\!11$.
        The mode content of the target harmonics $(2,2)$ and $(3,2)$ is determined correctly;
        the mode content of the $(4,2)$ harmonic is not reliable due to the absence of the $(5,2)$ harmonic and these rows are shaded in gray.
    }
    \begin{ruledtabular}
        \begin{tabular}{llllll}
            Iteration & Mode Added $\alpha'$ & Significance $\mathcal{S}_{\alpha'}$ & Mismatch $\mathcal{M}_{(2,2)}$ & Mismatch $\mathcal{M}_{(3,2)}$ & Mismatch $\mathcal{M}_{(4,2)}$ \\
            \hline
            1 & $(2, 2, 0, +)$ & $1 - \exp(-1.16\times 10^{7}) \approx 1$ & 0.00175 & 0.362 & 0.57 \\ %
            2 & $(3, 2, 0, +)$ & $1 - \exp(-1.31\times 10^{6}) \approx 1$ & 0.00151 & 0.00318 & 0.20 \\ %
            \rowcolor{gray!20}
            \multicolumn{6}{>{\columncolor{gray!20}}l}{} \\[-2.5ex]
            3 & $(4, 2, 0, +)$ & $1 - \exp(-1.90\times 10^{5}) \approx 1$ & 0.00151 & 0.00284 & 0.00471 \\ %
            4 & $(2, 2, 1, +)$ & $1 - \exp(-3.92\times 10^{4}) \approx 1$ & $1.49\times 10^{-5}$ & 0.00114 & 0.00401 \\ %
            5 & $(3, 2, 1, +)$ & $1 - \exp(-6.05\times 10^{3}) \approx 1$ & $5.46\times 10^{-6}$ & $8.69\times 10^{-5}$ & 0.00219 \\ %
            \rowcolor{gray!20}
            \multicolumn{6}{>{\columncolor{gray!20}}l}{} \\[-2.5ex]
            6 & $(4, 2, 1, +)$ & $1 - \exp(-2.35\times 10^{3}) \approx 1$ & $5.24\times 10^{-6}$ & $8.38\times 10^{-6}$ & 0.000358 \\ %
            \rowcolor{gray!20}
            \multicolumn{6}{>{\columncolor{gray!20}}l}{} \\[-2.5ex]
            7 & $(4, 2, 2, +)$ & $1 - \exp(-1.41\times 10^{2}) \approx 1$ & $5.26\times 10^{-6}$ & $6.82\times 10^{-6}$ & 0.000238 \\ %
            \rowcolor{gray!20}
            \multicolumn{6}{>{\columncolor{gray!20}}l}{} \\[-2.5ex]
            8 & $(4, 2, 3, +)$ & $1 - \exp(-1.11\times 10^{2}) \approx 1$ & $5.28\times 10^{-6}$ & $6.27\times 10^{-6}$ & 0.000137 \\ %
            9 & $(2, 2, 2, +)$ & $1 - \exp(-99.3)$ & $8.88\times 10^{-7}$ & $5.10\times 10^{-6}$ & 0.000137 \\ %
            \rowcolor{gray!20}
            \multicolumn{6}{>{\columncolor{gray!20}}l}{} \\[-2.5ex]
            10 & $(4, 2, 4, +)$ & $1 - \exp(-58.8)$ & $8.92\times 10^{-7}$ & $5.75\times 10^{-6}$ & $7.99\times 10^{-5}$ \\ %
            \rowcolor{gray!20}
            \multicolumn{6}{>{\columncolor{gray!20}}l}{} \\[-2.5ex]
            11 & $(4, 2, 5, +)$ & $1 - \exp(-24.0)$ & $8.87\times 10^{-7}$ & $5.26\times 10^{-6}$ & $5.90\times 10^{-5}$ \\ %
            \hline
            Stopping & \multicolumn{5}{l}{The next most significant mode $\alpha'=(2,2,3,+)$ has significance $\mathcal{S}_{\alpha'}=0.9997<0.9999$, below threshold.}
        \end{tabular}
    \end{ruledtabular}
    \caption{ \label{tab:spherical_harmonics}
        The spherical harmonics used for each numerical simulation. 
        For the non-precessing simulations 0001-0007 and 0010-0012 the harmonics satisfy $h_{\ell, m}=h_{\ell, -m}^*$, hence the $m<0$ harmonics are excluded. 
        For the equal-mass, non-precessing, identical-spin simulations 0001-0004 (and the `superkick' simulation 0009) the odd $m$ modes are suppressed and were excluded.  
        Target harmonics are indicated with an asterisk$^*$.
    }
    \begin{ruledtabular}
        \begin{tabular}{lr}
            Simulation catalog ID & Included spherical harmonics $\beta$\\
            \hline
            0001-0004 & $(2,2)^*, (3,2)^*, (4,2)$ \\
            0005-0007 & $(2,2^*), (3,3)^*, (3,2), (4,3)$ \\
            0008 & $(2,\pm 2)^*, (3,\pm 3)^*, (3,\pm 2), (4,\pm 3)$ \\
            0009 & $(2,\pm 2)^*, (3,\pm 2)^*, (4, \pm 2)$ \\ 
            0010-0012 & $(2,2)^*, (3,3)^*, (3,2), (4,4)^*, (4,3), (5,5)^*, (5,4), (6,6)^*, (6,5), (7,6)$  \\
            0013 & $(2,\pm 2)^*, (3,\pm 3)^*, (3,\pm 2), (4,\pm 4)^*, (4,\pm 3), (5,\pm 5)^*, (5,\pm 4), (6,\pm 6)^*, (6,\pm 5), (7,\pm 6)$ \\
            \end{tabular}
    \end{ruledtabular}
\end{table*}

\section{Power-Law Tail Analysis}

This section follows the notation introduced in the methods paper \cite{methods_paper} that describes the computationally efficient Bayesian methodology for analyzing the QNM content of the ringdown. 

The tail part of the model for the signal in the $\beta$ harmonic is given by 
\begin{align}
    T^\beta (t)=A^{\beta'}\delta^\beta_{\beta'}\left(t-t_{\beta'}\right)^{-\lambda_{\beta'}} ,\nonumber
\end{align}
where the Kronecker delta ensures the $\beta'$ tail only appears in one harmonic with no mode mixing.
This term is added to the model described in the main letter.
When quoting the amplitudes of the tails, we apply a decay correction to shift the amplitude to the desired start time analogous to that done for the QNM amplitudes $\hat{C}_\alpha$; the decay-corrected tail amplitude is defined as $\hat{A}^{\beta'} \!\equiv\! A^{\beta'} (t_0 \!-\! t_{\beta'})^{\lambda_{\beta'}}$.

We use a prior $\pi$ on the tail parameters which is flat in the real and imaginary parts of the amplitude in the ranges $-0.1<\mathrm{Re}A^{\beta'},\,\mathrm{Im}A^{\beta'}<0.1$, and flat on the reference time in the range $t_0 \!-\! 100M\!<\!t_{\beta'}\!<\!t_0 \!-\! 10M$. 
The power-law index $\lambda_{\beta'}$ is fixed to the predicted `Price' value; when fitting the Bondi news this is $\lambda_{\beta'} = 2 \ell + 2$. 
We also performed searches for tails where $\lambda_{\beta'}$ was included as a free parameter, with flat priors over the range $1 \leq \lambda_{\beta'} \leq 15$, and also saw no evidence for a tail.

The tail posterior, marginalized over the linear parameters $\vartheta$, is given by (sum over repeated $\mu,\,\nu$ indices) 
\begin{align}
    &\log P = \log \pi -\frac{1}{2}\sum_\beta \left<\mathfrak{h}^\beta-H_*^\beta-T^\beta|\mathfrak{h}^\beta-H_*^\beta-T^\beta\right>_\beta \nonumber \\
    &+\frac{1}{2} \left<h^\beta_\mu|\mathfrak{h}^\beta-H_*^\beta-T^\beta \right>_\beta (\Gamma^{-1})^{\mu\nu} \left<\mathfrak{h}^\beta-H_*^\beta-T^\beta|h^\beta_\nu \right>_\beta. \nonumber 
\end{align}
Tails were searched for one at a time using the mode content $\mathcal{D}$ detected by the algorithm at $t_0\!=\!50M$. 

The three-dimensional tail posterior was sampled using the \texttt{emcee} sampler \cite{2013PASP..125..306F} using 50 walkers evolved for 5000 iterations, which after removing 30\% of the chain as a burn in and thinning by twice the estimated autocorrelation length gave $\sim\!10^3$ independent posterior samples in the five posteriors for the tails in the target harmonics $\beta\!=\!(2,2), (3,3) \ldots (6,6)$. 

\end{document}